\documentclass[12pt]{article}
\usepackage{epsf}
\begin{document}

\begin{center}
{\bfseries SUPERNARROW DIBARYONS PRODUCTION IN $pd$
INTERACTIONS}

\vskip 5mm

L.V. Fil'kov$^{1 \dag}$,  
V.L. Kashevarov$^1$, E.S. Konobeevski$^{2 \dag}$, M.V. Mordovskoy$^2$, 
S.I. Potashev$^2$, V.A. Simonov$^2$, V.M. Skorkin$^2$, and S.V. Zuev$^2$

\vskip 5mm

{\small
(1) {\it Lebedev Physical Institute, Moscow, Russia}\\
(2) {\it Institute for Nuclear Research, Moscow, Russia}\\
$\dag$ {\it E-mail: filkov@x4u.lpi.ruhep.ru;\\ 
konobeev@sci.lebedev.ru}}
\end{center}

\vskip 5mm

\begin{center}
\begin{minipage}{150mm}
\centerline{\bf Abstract}
An analysis of new experimental data, obtained at the Proton Linear 
Accelerator of INR, is carried out with the aim of searching for
supernarrow dibaryons in the $pd\to p+pX_1$ and 
$pd\to p+dX_2$ reactions. Dibaryons with
masses 1904$\pm 2$, 1926$\pm 2$, and 1942$\pm 2$ MeV have been observed 
in invariant mass $M_{pX_1}$ spectra.
In missing mass $M_{X_1}$ spectra, the peaks  
at $M_{X_1}=966\pm 2$, 986$\pm 2$, and 1003$\pm 2$ MeV 
have been found. The analysis of the data obtained leads to the 
conclusion that the observed dibaryons are supernarrow dibaryons, the
decay of which into two nucleons is forbidden by the Pauli exclusion 
principle. A possible interpretation of "exited nucleon states" with small
masses is suggested. 
\end{minipage}
\end{center}

\vskip 10mm

The experimental search for dibaryons is continuing over 20 years
(see for review \cite{troy,tat1}). 
Usually one looks for NN coupled dibaryons. Such
dibaryons have decay widths from a few up to hundred MeV. Their
relative contributions are small enough but the background contribution
is big and uncertain as a rule. All this leads often to contradictory
results. 

We consider a new class of dibaryons - supernarrow dibaryons 
(SNDs), the decay of which into two nucleons
is forbidden by the Pauli exclusion principle
\cite{mul,fil1,fil2}. Such dibaryons with the mass
$M<2m_N+m_{\pi}$ ($m_N$($m_{\pi}$) is the nucleon (pion) mass)
can decay into two nucleons, mainly emitting a photon. Decay widths
of these dibaryons are $\le 1$ keV \cite{fil2}.

In Ref.\cite{izv,ksf,yad,prc}, we studied the reaction $pd\to pX$ 
with the aim of searching for supernarrow dibaryons.
The experiment was 
carried out at the proton beam of the Linear Accelerator of INR using
the two-arm spectrometer TAMS. As was shown in Ref. \cite{yad,prc},
the nucleons and the deuteron from the decay of SND into $\gamma NN$
and $\gamma d$ have to be emitted in a narrow angle cone with respect to
the direction of motion of the dibaryon. On the other hand, if a dibaryon
decays mainly into two nucleons, then the expected  angular cone of
emitted nucleons must be more than $50^{\circ}$. Therefore, a detection
of the scattered proton in coincidence with the proton (or the deuteron)
from the decay of particle $X$ at correlated angles allowed to suppress
essentially the contribution of the background processes and to increase
the relative contribution of a possible SND production. As a result,
two narrow peaks in missing mass spectra have been observed at
$M=$1905 and 1924 MeV with widths equal to the experimental resolution
(3 MeV). The analysis of the angular distributions of the charged particles
($p$ or $d$)
from the decay of particle $X$ showed that the peak found at 1905 MeV
most likely corresponds to a SND with isotopic spin equal to 1. 
In Ref. \cite{prc} arguments were presented for the resonance at
$M=$1924 MeV is a SND, too.

In order to argue more convincingly that the states found are really SNDs,
an additional experimental investigation was carried out.

\begin{figure}
\epsfxsize=14cm
\epsfysize=18cm
\centerline{
\epsfbox{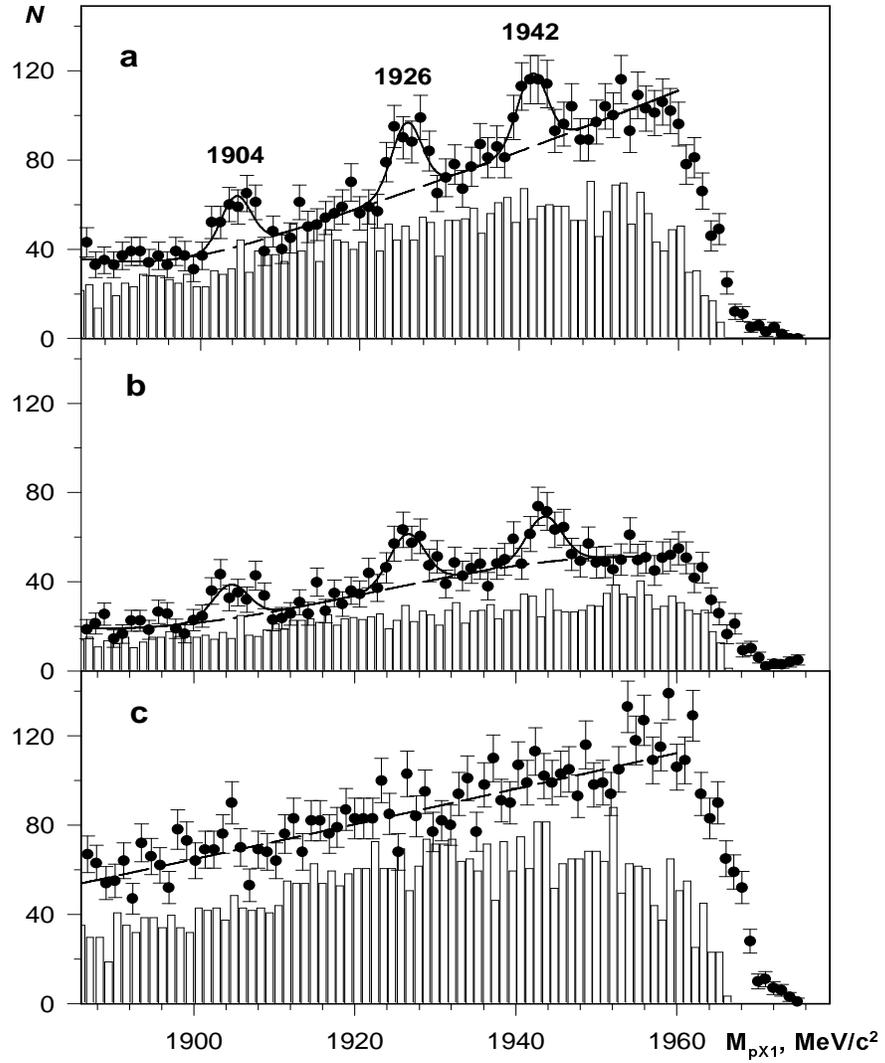}}
\caption{The invariant mass $M_{pX_1}$ spectra 
obtained with CD$_2$ (the points with the statistical errors) and $^{12}$C
(the bars) targets; 
(a) -- $\theta_R=34^{\circ}$, (b) -- $\theta_R=36^{\circ}$, 
(c) -- $\theta_R=38^{\circ}$. }
\end{figure} 

In the present paper we give the results of an analysis of
new experimental data respecting study of the 
$pd\to p+pX_1$ and $p\to p+dX_2$ processes 
at the Linear Accelerator of INR with 305 MeV proton beam using the 
spectrometer TAMS. The properties of this spectrometer are described
elsewhere \cite{prc}.
CD$_2$ and $^{12}$C were used as targets. 

In this experiment, the scattered proton was detected in the left arm
of the spectrometer TAMS at the angle $\theta_L=70^{\circ}$. The second
charged particle (either $p$ or $d$) was detected in 
the right arm by three telescopes located at $\theta_R=34^{\circ}$,
$36^{\circ}$, and $38^{\circ}$. These angles were shifted by $1^{\circ}$
in comparison with the conditions of our previous experiment,
that allowed to get an additional information about an angular distribution
of the charged particles from the decay of the dibaryons 
in question.

The angle $\theta_R=38^{\circ}$ corresponds to the elastic $pd$
scattering. A measurement of this effect was used to calibrate
the spectrometer.

At first, let us consider the reaction $pd\to p+pX_1$.

Figs. 1(a)-1(c) demonstrate
the experimental invariant mass $M_{pX_1}$ spectra 
obtained with the CD$_2$ (the points with the statistical
errors) and $^{12}$C (the bars) targets,
where (a), (b), and (c) correspond to a detection of the second
proton in the right arm detector at
$\theta_R=34^{\circ}$, $36^{\circ}$, and $38^{\circ}$, respectively.
Three peaks at $M_{pX_1}=1904\pm 2$, 1926$\pm 2$, and 1942$\pm 2$ MeV are 
observed in these spectra.
The first two of them confirmed the values of the dibaryon mass obtained
by us earlier \cite{izv,ksf,yad,prc} and the resonance at 1942 MeV is a
new one.  

The experimental invariant mass spectra obtained with the carbon target
are rather smooth. This 
smoothness is caused by both an essential increase of the contribution of
background reactions in the interaction of the proton with the carbon
and Fermi motion of nucleons in the nucleus. The latter increases
essentially the angular cone size of emitted nucleons. In consequence,
it is not possible to see peaks of SNDs in the present experiment 
on the carbon target.

As the experiment with
the carbon target resulted in the rather smooth spectra, all structures,
appearing in the experiment with the CD$_2$ target, may be explained
by an interaction of the proton with the deuteron. 

The calculation for the isovector SND $D(T=1,J^P=1^{\pm})$ with the
mass $M$=1904 MeV
showed that the contributions of such a dibaryon to spectra at the
angles $34^{\circ}$, $36^{\circ}$, and $38^{\circ}$ must
relate as $1:0.92:0.42$. For the isoscalar SND $D(0,0^{\pm})$ we obtained 
$1:0.95:0.67$.

The biggest contribution of the SND with $M=1926$ MeV 
is expected at $\theta_R=30^{\circ}-34^{\circ}$. The angles 
$30^{\circ}-33^{\circ}$ were not
investigated in this work. The calculation of the ratio of the contributions 
to the invariant mass spectra at the angles
$34^{\circ}$, $36^{\circ}$, and $38^{\circ}$ gave $1:0.85:0.34$ for $T=1$
and $1:0.71:0.46$ for $T=0$.

Nucleons from the decay of the SND with $M=1942$ MeV must 
have a wider angular 
distribution with maximum in the region of $26^{\circ}-32^{\circ}$.
In the region of the angles under consideration in this work,  
the contributions of $D(1,1^{\pm})$ are expected to be in the ratio of 
$1:0.6:0.2$. For the SND $D(0,0^{\pm})$ we have $1:0.78:0.55$.

\begin{figure}
\epsfxsize=14cm
\epsfysize=10cm
\centerline{
\epsfbox{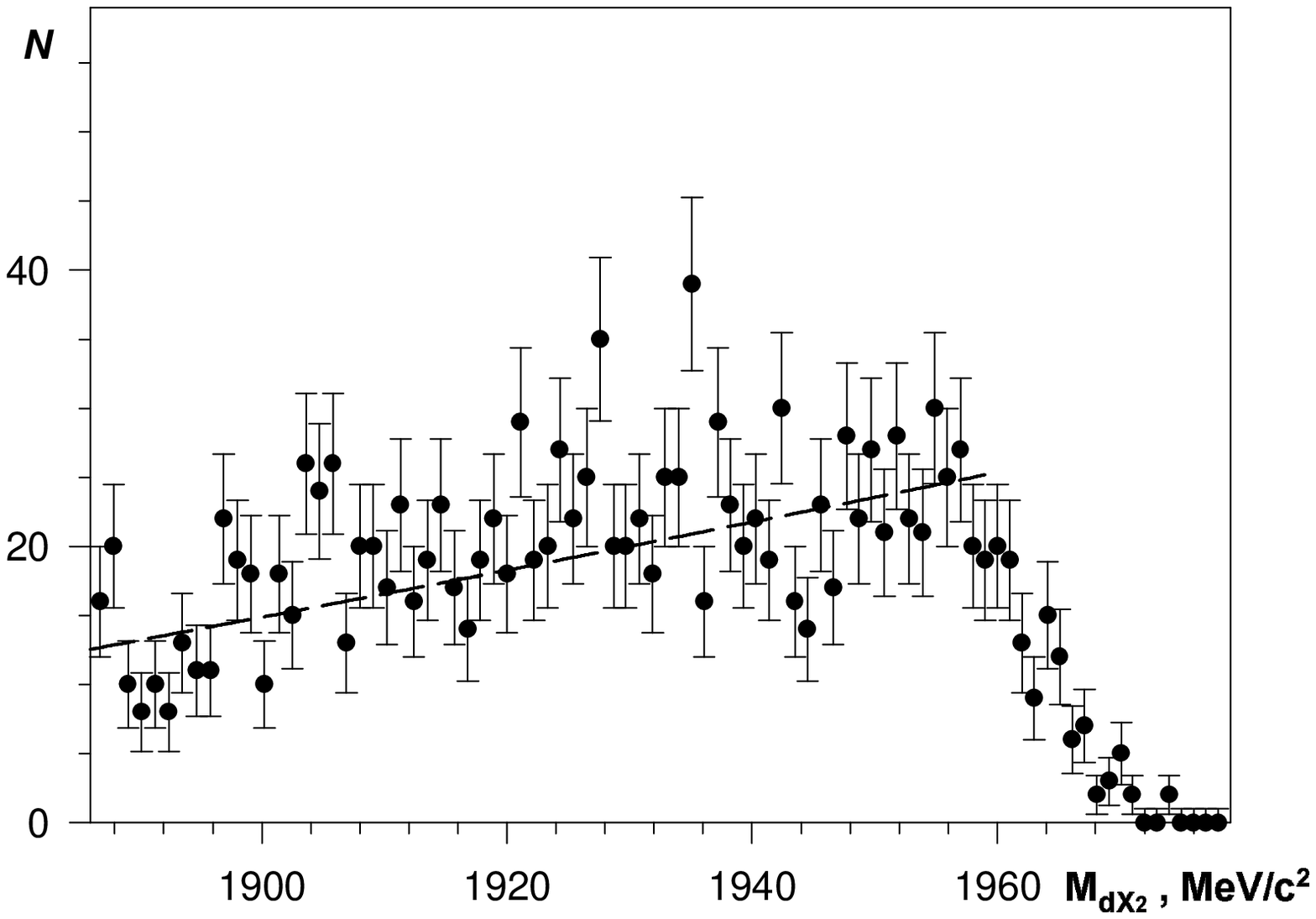}}
\caption{The invariant mass $M_{dX_2}$ spectra 
for the reaction $pd\to p+dX_2$ for the sum of the angles
$\theta_R=34^{\circ}$ and $36^{\circ}.$} 
\end{figure} 

All these predictions for the SNDs are in agreement with our 
experimental data within the errors. However, the analysis of
the reaction $pd\to p+pX_1$ only in the considered angle range  
does not allow to determine an isotopic spin of the SNDs.

If the observed states are $NN$-coupled dibaryons decaying 
mainly into two nucleons then the expected angular
cone size of emitted nucleons must be more than $50^{\circ}$.
Therefore, their contributions to the invariant mass spectra in 
Fig. 1(a)-1(b), would
be nearly the same and would not exceed a few events, even assuming that
the dibaryon production cross section is equal to that of elastic $pd$
scattering ($\sim 40 \mu$b/sr). Hence, the peaks found
most likely correspond to SNDs.

The invariant mass spectrum $M_{dX_2}$ of the reaction $pd\to p+dX_2$,
for the sum of
angles $\theta_R=34^{\circ}$ and $36^{\circ}$, is shown in
Fig. 2. As seen from this figure, the reaction $pd\to p+dX_2$ gives very 
small contribution into the production of the dibaryons under study.

On the other hand, it is expected \cite{yad,prc} that isoscalar SNDs 
contribute mainly
into $\gamma d$ channel and isovector SNDs do into $\gamma NN$ one.
As the main contribution of the found dibaryons is observed in $pX_1$
channel, it is possible to expect that $X_1=\gamma +n$, and
this is an indication that the all found states are isovector SNDs.
The more precise conclusion about the value of the  isotopic spin of the 
observed SNDs could be obtained by the study of the reaction
$pd\to n+X$. 

\begin{figure}
\epsfxsize=14cm
\epsfysize=16cm
\centerline{
\epsfbox{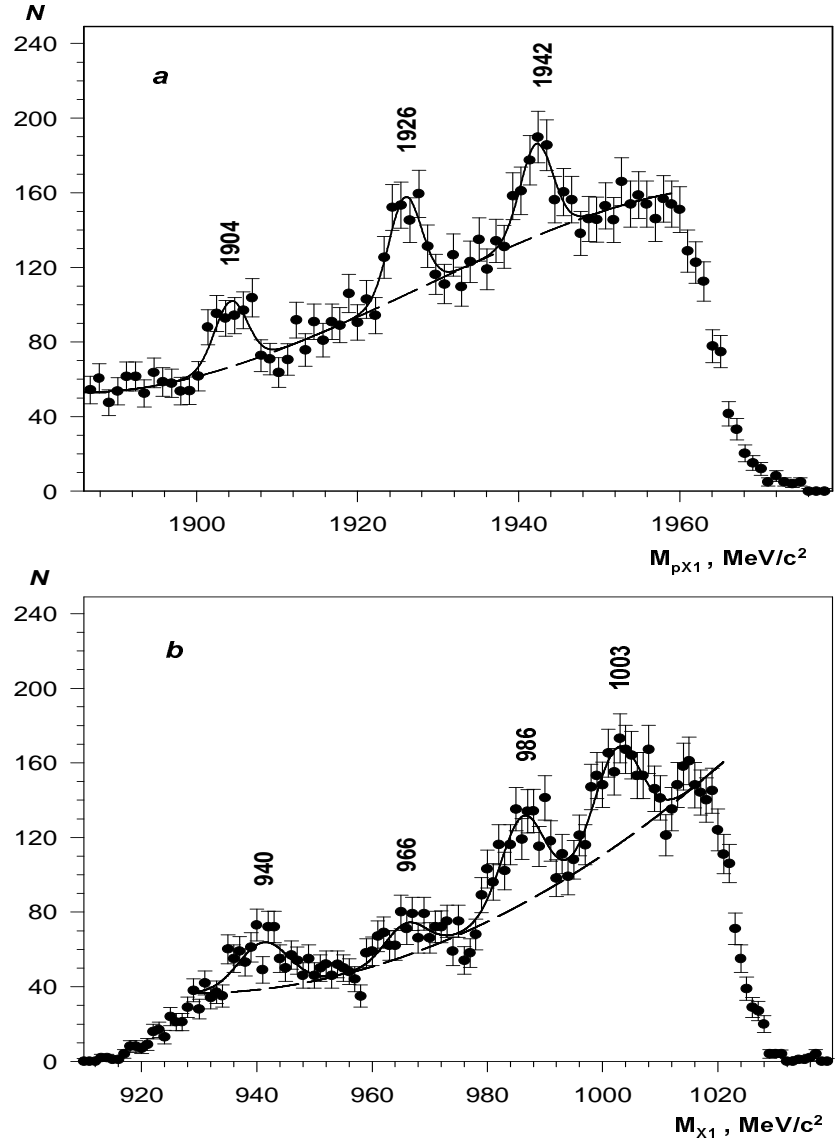}}
\caption{The invariant mass $M_{pX_1}$ (a) and the missing mass 
$M_{X_1}$ (b) spectra for the sum of angles of
$\theta_R=34^{\circ}$ and $\theta_R=36^{\circ}$.}
\end{figure} 

The summary spectrum of the reaction $pd\to p+pX_1$ over 
angles $\theta_R=34^{\circ}$ and $36^{\circ}$, where the contribution of 
the SNDs is maximum,
is presented in Fig. 3(a). This spectrum was interpolated by a
second order polynomial (for the background) plus Gaussians (for the
peaks). The number of standard deviations (SD) is determined as
$$
\frac{N_{eff}}{\sqrt{N_{eff}+N_{back}}}
$$
where $N_{eff}$ is the number of events above the background curve and
$N_{back}$ is the number of events below this curve.
Taking nine points for each peak, we have 6.0, 7.0, and 6.3 SD
for the resonances at 1904, 1926, and 1942 MeV, respectively. The widths of 
these resonances are equal to the experimental resolution of $\sim 4$ MeV.

An additional information about the nature of the observed states 
was obtained by studying the missing mass $M_{X_1}$ spectra of the
reaction $pd\to p+pX_1$.
If the state found is a dibaryon decaying mainly into two nucleons then
$X_1$ is a neutron and the mass $M_{X_1}$ is equal to the neutron mass
$m_n$. If the value of $M_{X_1}$, obtained from the experiment, differs
essentially from $m_n$ then $X_1=\gamma+n$ and we have the additional 
indication that the observed dibaryon is SND.

The simulation of missing mass spectra for the reaction $pd\to p+pX_1$, 
where $pX_1$ are decay products of the SNDs 
with the masses 1904, 1926, and 1942 MeV, gave peaks
at $M_{X_1}$=965, 987, and 1003 MeV, respectively.

Fig. 3(b) demonstrates the missing mass $M_{X_1}$ spectrum 
obtained from the experiment 
for the sum of the angles $\theta_R=34^{\circ}$ and $36^{\circ}$.
As is seen from this figure, besides the peak at neutron mass, 
which caused by the process $pd\to p+pn$,
a resonancelike behavior of the spectrum is observed at $966\pm 2$,
$986\pm 2$, and $1003\pm 2$ MeV. These values of $M_{X_1}$ coincide with
the ones obtained  from the simulation and differ essentially from 
the value of the neutron mass (939.6 MeV). Hence, for all states under
study, $X_1=\gamma+n$ and the dibaryons found with the masses 1904,
1926, and 1942 MeV are really SNDs.

It should be noted that the peak at
$M_{X_1}=1003\pm 2$ MeV corresponds to the resonance found in \cite{tat2}
and attributed to an excited nucleon state $N^*$. 
In this work, the authors brought out
three such states with masses 1004, 1044, and 1094 MeV. 
In principle, SND could decay into $N N^*$. A possibility of the
production of $NN^*$-coupled dibaryons was considered in \cite{tat1}.

If these excited
nucleon states decay into $\gamma N$ then they would contribute
to the Compton scattering on the nucleon. However, the analysis 
\cite{lvov} of the experimental data on this process completely excludes 
$N^*$ as intermediate state in the Compton scattering on the nucleon.
 
In Ref. \cite{kob} it was assumed that these states belong to totally
antisymmetric $\underline{20}$-plet of the spin-flavor $SU(6)_{FS}$ group. 
Such a $N^*$ can transit into 
nucleon only if two quarks from the $N^*$ participate in the
interaction \cite{feyn}. Then the simplest decay of $N^*$ with the masses
1004 and 1044 MeV would be $N^*\to \gamma\gamma N$.
This assumption could be checked, in particular,
by studying the reactions $\gamma p\to\gamma X$ or $\gamma p\to\pi X$
at the photon energy close to 800 MeV.

Taking into account the found connection between the SNDs 
and the resonancelike states $X_1$, 
it is possible to assume that the peaks, observed in 
\cite{tat2} at 1004 and 1044 MeV, are not excited nucleons, but
they are the resonancelike states $X_1=\gamma+n$ 
caused by possible existence and decay of the SNDs with the masses 1942 
and 1982 MeV, respectively. 
Such $X_1$ are not real resonances and cannot give contribution to the
Compton scattering on the nucleon. 


The following conclusion can be made. As a result of the study of the
reaction $pd\to p+pX_1$ three narrow peaks at 1904, 1926,
and 1942 MeV have been observed in the invariant mass $M_{pX_1}$ spectra.
The analysis of the angular distributions of the protons from decay of
$pX_1$ states and the data on the reaction $pd\to p+dX_2$
showed that the peaks found can be explained as a 
manifestation of the isovector SNDs, the decay of which into two 
nucleons is forbidden by the Pauli exclusion principle.
The observation of the peaks in the missing mass
$M_{X_1}$ spectra at 966, 985, and 1003 MeV is an additional confirmation
that the dibaryons found are the SNDs.

The authors thank L.V. Kravchuk and V.M. Lobashev for support of this
work and the team of the Linear Accelerator for the help in performance
of the experiment. The authors thank also B.M. Ovchinikov and 
J. Friedrich for the active discussion of the experimental results.

\end{document}